\documentclass[cite]{epl}
\usepackage{amsmath}
\usepackage{graphicx}
\usepackage{amsfonts}
\usepackage{amssymb}
\shorttitle{Controlling spin...}
\institute{
  Department of Physics and Astronomy, University of Basel,
  Klingelbergstrasse~82,
  CH-4056 Basel, Switzerland}
\pacs{73.23.-b}{Electronic transport in mesoscopic systems}
\pacs{75.75.+a}{Magnetic properties of nanostructures}
\pacs{85.75.-d}{Magnetoelectronics; spintronics: devices exploiting spin polarized transport or integrated magnetic fields}

\begin{document}

\title{Controlling spin in an electronic interferometer with spin-active interfaces}
\author{A. Cottet, T. Kontos, W. Belzig, C. Sch\"{o}nenberger and C. Bruder}
\maketitle
\begin{abstract}
We consider electronic current transport through a ballistic one-dimensional
quantum wire connected to two ferromagnetic leads. We study the effects of the
\textit{spin-dependence} of interfacial phase shifts (SDIPS) acquired by
electrons upon scattering at the boundaries of the wire. The SDIPS produces a
spin splitting of the wire resonant energies which is tunable with the gate
voltage and the angle between the ferromagnetic polarizations. This property
could be used for manipulating spins. In particular, it leads to a giant
magnetoresistance effect with a sign tunable with the gate voltage and the
magnetic field applied to the wire.
\end{abstract}

The quantum mechanical spin degree of freedom is now widely exploited to
control current transport in electronic devices \cite{Prinz}. However, one
major functionality to be explored is the electric field control of spin. In
the context of a future spin electronics or spintronics, this would allow to
build the counterpart of the field-effect transistor (FET), namely the
spin-FET, in which spin transport would be controlled through an electrostatic
gate \cite{Datta,Schapers}. In devices where single spins are used to encode
quantum information, this property should also allow to perform single quantum
bit operations by using effective magnetic fields which would be locally
controllable with the gate electrostatic potential \cite{LossDiVicenzo}. Among
the potential candidates for implementing the electric field control of the
spin dynamics, spin-orbit coupling seems a natural choice \cite{Datta}.
However, whether it is possible to use spin-orbit coupling to make spin-FETs
or spin quantum bits is still an open question \cite{Zutic}.
\begin{figure}[ptb]
\centering\includegraphics[width=0.3\linewidth]{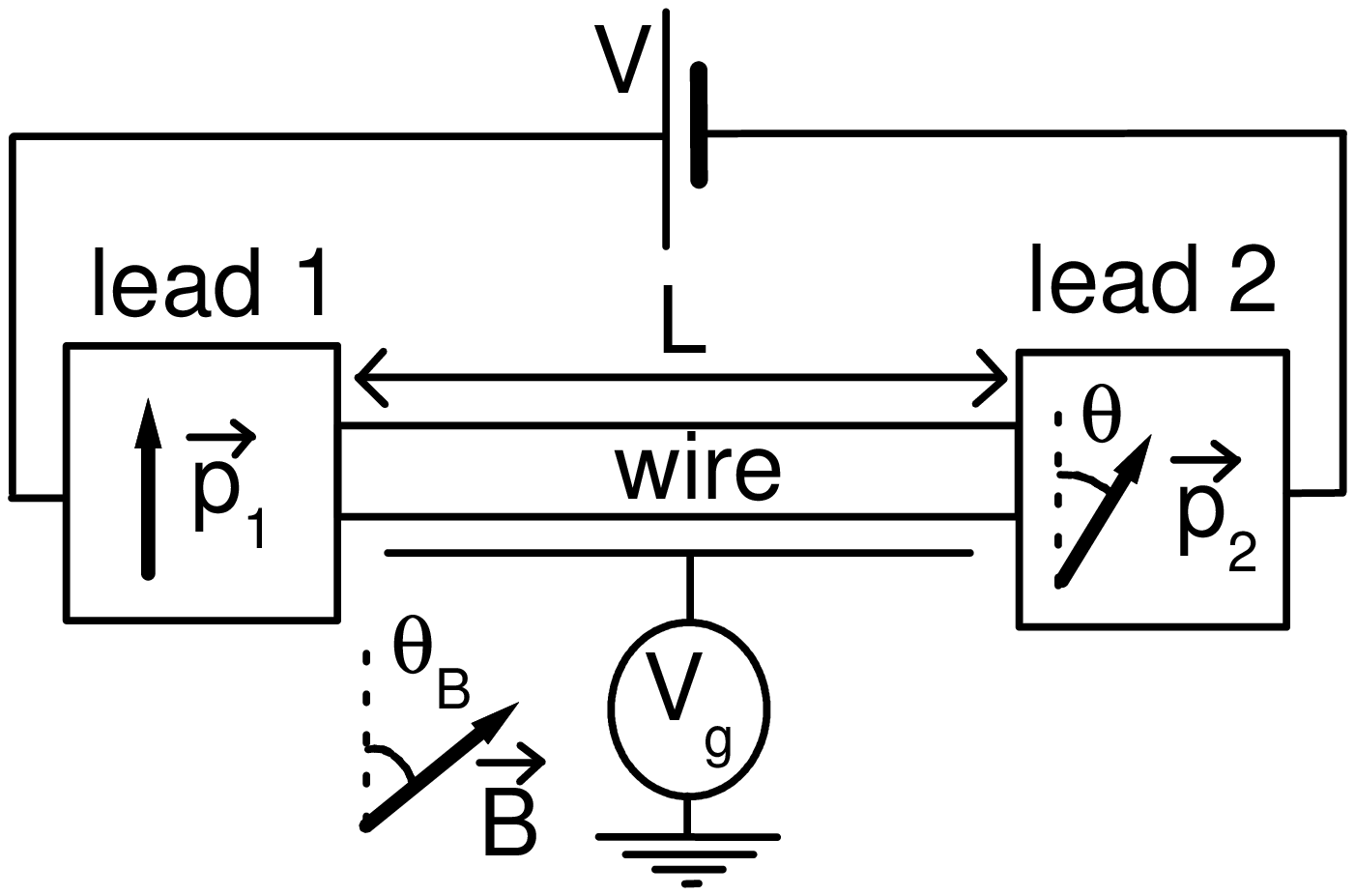}\qquad
\includegraphics[width=0.6\linewidth]{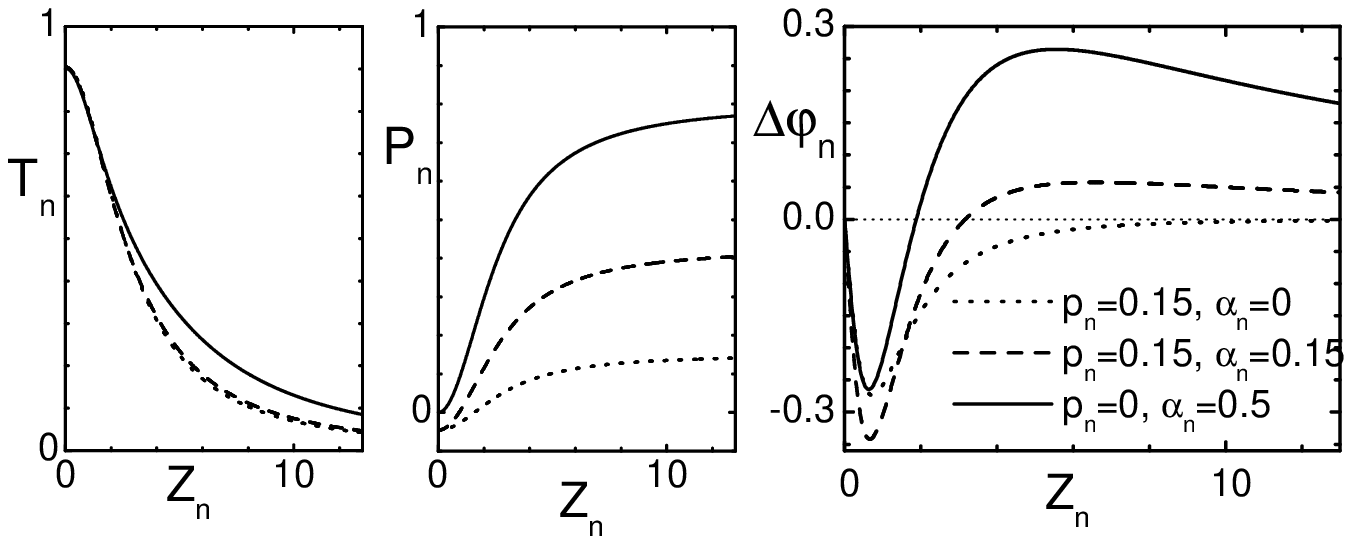}\caption{Left: Electrical
diagram of a ballistic wire of length $L$ connected to ferromagnetic leads $1$
and $2$ with polarizations $\vec{p}_{1}$ and $\vec{p}_{2}$. The wire is
capacitively coupled to a gate voltage source $V_{g}$. A magnetic field
$\vec{B}$ is applied to the circuit. We assume that $\vec{p}_{1}$, $\vec
{p}_{2}$ and $\vec{B}$ are coplanar, with angles $\theta=(\widehat{\vec{p}%
_{1},\vec{p}_{2}})$ and $\theta_{B}=(\widehat{\vec{p}_{1},\vec{B}})$. Right:
Spin-averaged tunneling rate $T_{n}$ (left panel), tunneling rate polarization
$P_{n}$ (middle panel) and SDIPS parameter $\Delta\varphi_{n}$ (right panel)
of contact $n\in\{1,2\}$, estimated by using a Dirac barrier with a
spin-dependent coefficient $U_{n}^{s}$ (see \cite{BTK}), placed between a
ferromagnetic metal with Fermi energy $E_{F}^{n}=10~\mathrm{eV}$, and a wire
with Fermi wavevector $k_{Fw}=8.5~10^{9}\mathrm{m}^{-1}$ typical of single
wall nanotubes \cite{bockrath}. We show the results as a function of the
average barrier impedance $Z_{n}=m_{e}(U_{n}^{u}+U_{n}^{d})/(\hbar^{2}k_{Fw}%
)$, for different values of the polarization $p_{n}$ of lead $n$ and of the
spin asymmetry $\alpha_{n}=(U_{n}^{\downarrow}-U_{n}^{\uparrow})/(U_{n}%
^{\uparrow}+U_{n}^{\downarrow})$ of the barrier. When the barrier is
considered to be spin-independent, i.e. $\alpha_{n}=(U_{n}^{\downarrow}%
-U_{n}^{\uparrow})/(U_{n}^{\uparrow}+U_{n}^{\downarrow})=0$, $\Delta
\varphi_{n}$ can be finite for $T_{n}$ large only. (Here the wave-vector
mismatch between the lead and the wire leads to $\Delta\varphi_{n}<0$ and can
also\ lead to $P_{n}<0$ for $Z_{n}$ small). It is also possible to assume
$\alpha_{n}>0$ i.e., the barrier is magnetically polarized, with the same
polarization direction as lead $n$. This can be caused by the magnetic
properties of the contact material, but it can also be obtained artificially
by using a magnetic insulator to form the barrier. In this case a large
$\Delta\varphi_{n}$ can be obtained for $T_{n}$ small also, with
$\Delta\varphi_{n}>0$ due to a weaker penetration of minority electrons in the
barrier. This shows that it is relevant to study the effect of the SDIPS (i.e.
having $\Delta\varphi_{n}\neq0$) for a wide range of $T_{n}$.}%
\label{Figure1}%
\end{figure}

In this context, it is crucial to take into account that the interface between
a ferromagnet and a non-magnetic material can scatter electrons with spin
parallel or antiparallel to the magnetization of the ferromagnet with
different phase shifts. This \textit{spin-dependence} of interfacial phase
shifts (SDIPS) can modify significantly the behavior of hybrid circuits.
First, the SDIPS implies that spins non-collinear to the magnetization of the
ferromagnet precess during the interfacial scattering, like the polarization
of light rotates upon crossing a birefringent medium. This precession is
expected to increase the current through diffusive F/normal metal/F spin
valves when the magnetizations of the two F electrodes are non-collinear
\cite{FNF}. The same phenomenon is predicted to occur in F/Luttinger liquid/F
\cite{Luttinger} and F/Coulomb blockade island/F\cite{Wetzels} spin valves in
the incoherent limit. In collinear configurations, precession effects are not
relevant, but the SDIPS can affect mesoscopic coherence phenomena. For
instance, in superconducting/F hybrid systems \cite{Tokuyasu,otherBC,SF2005},
the SDIPS manifests itself by introducing a phase shift between electrons and
holes coupled coherently by Andreev reflections. References \cite{Tokuyasu}
and \cite{SF2005} have identified experimental signatures of this effect in
the data of \cite{Tedrow} and \cite{Takis}, respectively. However, the SDIPS
had not been shown to affect normal systems in collinear configurations up to now.

In this Letter, we show that the SDIPS can indeed affect normal systems in
collinear configurations, leading to properties which could be used for
controlling spins in the context of spintronics and quantum computing. We
consider a non-interacting one-dimensional ballistic wire contacted by two
ferromagnetic leads. In this system, Fabry-Perot-like energy resonances occur
due to size quantization, as observed experimentally for instance in carbon
nanotubes contacted by normal electrodes \cite{bockrath}. We show that the
SDIPS modifies qualitatively the behavior of this device even in the collinear
configuration, due to coherent multiple reflections. More precisely, we
explain how the SDIPS leads to a spin-splitting of the resonant energies which
is tunable with the gate voltage and the angle between the ferromagnetic
polarizations. This provides a justification for an heuristic approach which
was introduced recently by three of us for fitting magnetoresistance data
obtained in single wall nanotubes connected to ferromagnetic leads in the
collinear configuration \cite{Sahoo}. In this experiment, the estimated SDIPS
was relatively weak. We show that a larger SDIPS could be obtained by
engineering properly the contacts to the wire, and that this could be used for
manipulating the spin degree of freedom. In particular, the SDIPS-induced
spin-splitting could lead to a giant magnetoresistance effect with a sign
tunable with the gate voltage and the magnetic field, which should allow to
build very efficient spin-FETs.

We consider a single-channel ballistic wire of length $L$ contacted by
ferromagnetic leads $1$ and $2$ (Fig. \ref{Figure1}). This wire is
capacitively coupled to a gate biased at a voltage $V_{g}$, which allows to
tune its chemical potential. The directions of the magnetic polarizations
$\vec{p}_{1}$ and $\vec{p}_{2}$ of leads $1$ and $2$ form an angle
$\theta=(\widehat{\vec{p}_{1},\vec{p}_{2}})$. The spin states parallel
(antiparallel) to $\vec{p}_{n}$ are denoted $\uparrow_{n}(\downarrow_{n})$ in
general expressions, or $u(d)$ in expressions referring explicitly to lead $n$
only. The wire is subject to a DC magnetic field $\vec{B}$ coplanar to
$\vec{p}_{1} $ and $\vec{p}_{2}$, with $\theta_{B}=(\widehat{\vec{p}_{1}%
,\vec{B}})$. Following \cite{Veillette}, we use a scattering description
\cite{Blanter} in which an interface L/R is described by a scattering matrix
$S$ such that $[\widehat{a}_{L,-}^{\uparrow_{1}},\widehat{a}_{R,+}%
^{\uparrow_{1}},\widehat{a}_{L,-}^{\downarrow_{1}},\widehat{a}_{R,+}%
^{\downarrow_{1}}]^{t}=S[\widehat{a}_{L,+}^{\uparrow_{1}},\widehat{a}%
_{R,-}^{\uparrow_{1}},\widehat{a}_{L,+}^{\downarrow_{1}},\widehat{a}%
_{R,-}^{\downarrow_{1}}]^{t}$, with $\widehat{a}_{L[R],+\{-\}}^{s}$ the
annihilation operator associated to the right-going (left-going) electronic
state with spin $s$ at the left[right] side of the interface (we use spin
space $\{\uparrow_{1},\downarrow_{1}\}$ for defining $S$). In the low $V$
limit, the electrostatic potential of the wire is $\alpha V_{g}$, with
$\alpha$ the ratio between the gate capacitance and the total wire
capacitance. We assume that $e\alpha V_{g},g\mu_{B}B\ll E_{Fw}$, with $E_{Fw}$
the Fermi energy of the wire, $g$ the Land\'{e} factor, $\mu_{B}$ the Bohr
magneton and $e>0$ the electronic charge. Then, the propagation of electrons
with energy $E\simeq E_{Fw}$ through the wire is described by a scattering
matrix $S_{w}=P(\theta_{B})\left(  \exp\left[  i\delta\sigma^{0}-i\sigma
^{z}(\gamma_{B}/2)\right]  \otimes\tau_{1}\right)  P^{-1}(\theta_{B})$ with
$\delta=L(k_{Fw}+[(E+e\alpha V_{g}-E_{Fw})/\hbar v_{Fw}])$, $\gamma_{B}%
=g\mu_{B}BL/\hbar v_{Fw}$, $P\left[  \vartheta\right]  =[\cos(\vartheta
/2)\sigma_{0}-i\sin(\vartheta/2)\sigma_{y}]\otimes\tau_{0}$, and $k_{Fw}$
($v_{Fw}$) the Fermi wave-vector (velocity) in the wire. Here, $\sigma_{i}$,
with $i\in\{0,x,y,z\}$, are the Pauli matrices acting on spin space
$\{\uparrow_{1},\downarrow_{1}\}$. The matrices $\tau_{i}$, with
$i\in\{0,1,2,3\}$, are the Pauli matrices relating the space of incoming
electrons $\{(L,+),(R,-)\}$ to the space of outgoing electrons
$\{(L,-),(R,+)\}$. We assume that there is no spin flip between the states
$\uparrow_{n}$ and $\downarrow_{n}$ while electrons tunnel through interface
$n$. Then, the scattering matrices describing the contacts $1$ and $2$ are
respectively $S_{1}=\tilde{S}_{1}$ and $S_{2}=P[\theta]\tilde{S}_{2}%
P^{-1}[\theta]$ with $2\tilde{S}_{n}=(\sigma_{0}+\sigma_{z})\otimes\tilde
{s}_{n}^{u}+(\sigma_{0}-\sigma_{z})\otimes\tilde{s}_{n}^{d}$ and
\begin{equation}
\tilde{s}_{n}^{s}=\left[
\begin{array}
[c]{cc}%
r_{n,L}^{s} & t_{n,R}^{s}\\
t_{n,L}^{s} & r_{n,R}^{s}%
\end{array}
\right]  \label{matrix}%
\end{equation}
for $s\in\{u,d\}$. Here, $t_{n,m}^{s}$ and $r_{n,m}^{s}$ are complex
amplitudes of transmission and reflection for electrons with spin $s$,
incident from the side $m\in\{L,R\}$ of contact $n\in\{1,2\}$. We also define
the transmission probability $T_{n}^{s}=|t_{n,L(R)}^{s}|^{2}=1-|r_{n,L(R)}%
^{s}|^{2}$. We assume that electron-electron interactions can be neglected.
Then, the linear conductance of the wire at temperature $T$ is $G=G_{Q}%
\sum\nolimits_{s\in\{\uparrow_{1},\downarrow_{1}\},r\in\mathbb{B}}%
\int\nolimits_{0}^{+\infty}\mathbb{T}_{sr}(E)(-\partial n_{F}(E)/\partial E)$,
with $G_{Q}=e^{2}/h$, $n_{F}(E)=1/[1+\exp(E/k_{B}T)]$ and $\mathbb{T}_{sr}$
the probability that an electron with spin $s$ from lead 1 is transmitted as
an electron with spin $r$ to lead 2 (we will use $\mathbb{B=}\{\uparrow
_{1},\downarrow_{1}\} $ or $\mathbb{B=}\{\uparrow_{2},\downarrow_{2}\}$,
depending on convenience, for describing the spin state $r$ in lead 2). The
transmissions $\mathbb{T}_{sr}$ can be found from the global scattering matrix
$S_{tot}=S_{1}\circ S_{w}\circ S_{2}$ associated to the device (see e.g. Ref.
\cite{Cahay} for the definition of the composition rule $\circ$). In the
configurations studied in this Letter, the only interfacial scattering phases
remaining in $\mathbb{T}_{sr}$ are the reflection phases at the side of the
wire, i.e. $\varphi_{1}^{u[d]}=\arg(r_{1,R}^{u[d]})$ and $\varphi_{2}%
^{u[d]}=\arg(r_{2,L}^{u[d]})$. Importantly, these phases depend on spin
because, due to the ferromagnetic exchange field, electrons are affected by a
spin-dependent scattering potential when they encounter the interface between
the wire and lead $n$. We will characterize this spin-dependence with the
SDIPS parameters $\Delta\varphi_{n}=\varphi_{n}^{u}-\varphi_{n}^{d}$, with
$n\in\{1,2\}$. We also define the average tunneling rate $T_{n}=(T_{n}%
^{u}+T_{n}^{d})/2$ and the tunneling rate polarization $P_{n}=(T_{n}^{u}%
-T_{n}^{d})/(T_{n}^{u}+T_{n}^{d})$. In principle, the parameters $T_{n}$,
$P_{n}$ and $\Delta\varphi_{n}$ depend on the microscopic details of barrier
$n$, but general trends can already be found from simple barrier models (see
e.g. Figure \ref{Figure1}). When there is a spin-independent barrier between
the wire and lead $n$, $\Delta\varphi_{n}$ can be finite for $T_{n}$ large
only because a strong barrier prevents reflected electrons from being affected
by the lead magnetic properties. However, it is likely that the barrier
between the wire and the lead is itself spin-polarized. This can be due to the
magnetic properties of the contact material, but it can also be obtained
artificially by using a magnetic insulator like e.g. EuS (see \cite{MI}) to
form the barrier. In this case a large $\Delta\varphi_{n}$ can be obtained for
$T_{n}$ small also (see e.g. full lines in Fig.1, right). It is thus relevant
to study the effects of the SDIPS (i.e. having $\Delta\varphi_{n}\neq0$) for a
wide range of $T_{n}$.

\begin{figure}[ptb]
\centering\includegraphics[width=0.6\linewidth]{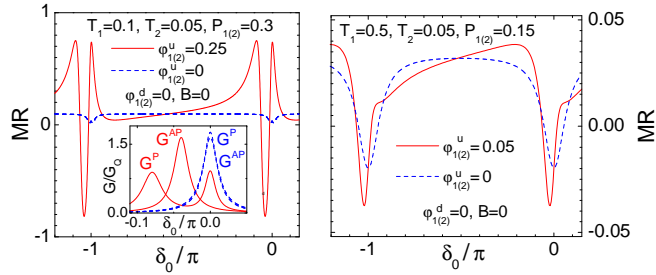}\caption{Left
panel: wire linear conductances $G^{P}=G(\theta=0)$ and $G^{AP}=G(\theta=\pi)$
(inset) and magnetoresistance $MR=(G^{P}-G^{AP})/(G^{P}+G^{AP})$ (main frame),
as a function of the spin-averaged phase $\delta_{0}$ acquired by electrons
upon crossing the wire ($\delta_{0}$ is linear with $V_{g}$ in the limit
studied in this paper). We used $T_{1}=0.1$, $T_{2}=0.05$, $P_{1}=P_{2}=0.3$,
$B=0$, and no SDIPS (blue dashed lines) or a SDIPS characterized by
$\varphi_{1}^{u}=\varphi_{2}^{u}=0.25$ and $\varphi_{1}^{d}=\varphi_{2}^{d}=0$
(red full lines). The SDIPS produces a spin splitting of the conductance peaks
because it shifts the phases accumulated by spins $\uparrow$ and $\downarrow$
upon reflection at the boundaries of the wire. This effect strongly increases
the magnetoresistance of the device. Right panel: Magnetoresistance $MR$ as a
function of $\delta_{0}$, for a device with $T_{1}=0.5$, $T_{2}=0.05$,
$P_{1}=P_{2}=0.15$, $B=0$ and no SDIPS (blue dashed lines) or a SDIPS such
that $\varphi_{1}^{u}=\varphi_{2}^{u}=0.05$ and $\varphi_{1}^{d}=\varphi
_{2}^{d}=0$ (red full lines). In this case, the spin-splitting of the
conductance curves cannot be resolved but the symmetry of the MR oscillations
is broken due to a conjugated effect of the SDIPS and the finite polarization
of the transmission probabilities.}%
\label{Figure3}%
\end{figure}

We now present the results given by the scattering description of the circuit.
We assume temperature $T=0$ and postpone a discussion on the effects of finite
temperatures to the end of this Letter. We first consider the case of parallel
($\theta=0$, noted $P$) or antiparallel ($\theta=\pi$, noted $AP$) lead
polarizations, with $\theta_{B}=0$. We note $\overline{\uparrow_{n}%
}=\downarrow_{n}$ and $\overline{\downarrow_{n}}=\uparrow_{n} $. In
configuration $c\in\{P,AP\}$, one has $\mathbb{T}_{s\overline{s}}^{c}=0$,
which means that spin is conserved when electrons cross the wire. The
conductance of the device can be calculated from $\mathbb{T}_{ss}^{c}%
=A_{ss}^{c}/|\mathbb{\beta}_{ss}^{c}|^{2}$, with $s\in\{\uparrow
_{1},\downarrow_{1}\}$, $A_{sr}^{P[AP]}=T_{1}^{s}T_{2}^{r[\overline{r}]}$ and
$B_{sr}^{P[AP]}=[(1-T_{1}^{s})(1-T_{2}^{r[\overline{r}]})]^{1/2}$. The term
\begin{equation}
\mathbb{\beta}_{sr}^{P[AP]}=1-B_{sr}^{P[AP]}e^{i(\varphi_{1}^{s}+\varphi
_{2}^{r[\overline{r}]}+2\delta+\kappa_{s}^{1}\gamma_{B})} \label{beta}%
\end{equation}
with $\kappa_{u(d)}^{n}=\pm1$ for $n\in\{1,2\}$, accounts for multiple
reflections between the two contacts (we have used indices $r\neq s$ and
$n\in\{1,2\}$ in the above formulas for later use). The transmission
probability $\mathbb{T}_{ss}^{P[AP]}$ for spins $s\in\{\uparrow_{1}%
,\downarrow_{1}\}$ is maximum at resonant energies $E_{s}^{P[AP],j}=(2\pi
j-\varphi_{1}^{s}-\varphi_{2}^{s[\overline{s}]}-\kappa_{s}^{1}\gamma
_{B})(\hbar v_{Fw}/2L)-e\alpha V_{g}-E_{Fw}$, with $j\in\mathbb{Z}$.
Importantly, these resonant energies depend on spin $s$ due to the SDIPS. This
leads to the conclusion that the SDIPS can modify the conductance of a normal
spin valve \textit{even in a collinear configuration}. This feature is due to
the ballistic nature of the system which offers the possibility of coherent
multiple reflections between the contacts. Note that from Eq.~(\ref{beta}),
the SDIPS affects electrons in the same way as a magnetic field collinear to
the lead polarizations. However, the spin-splitting induced by the SDIPS can
be different in the $P$ and the $AP$ configurations, contrarily to the
splitting produced by a magnetic field collinear to $\vec{p}_{1}$ and $\vec
{p}_{2}$. Indeed, in the $P[AP]$ configuration, there is a spin-splitting of
the resonant energies if $\Delta\varphi^{P[AP]}=\Delta\varphi_{1}%
+[-]\Delta\varphi_{2}\neq0$. In particular, the spin-splitting vanishes in the
$AP$ case when the contacts are perfectly symmetric. In the limit of low
transmissions $T_{n}^{u(d)}\ll1$, one can expand $\mathbb{T}_{ss}^{c}(E)$
around $E=E_{s}^{c,j}$ (see \cite{Blanter}) to derive the Breit-Wigner formula
$\mathbb{T}_{ss}^{P[AP]}=A_{ss}^{P[AP]}[(2L[E-E_{s}^{P[AP],j}]/\hbar
v_{Fw})^{2}+(T_{1}^{s}+T_{2}^{s[\overline{s}]})^{2}/4]^{-1}$ introduced
heuristically in \cite{Sahoo}. This equation shows that the spin-splitting
$\Delta\varphi^{c}$ can be fully resolved in the conductance curve
$G^{c}(V_{g})$ associated to configuration $c=P[AP]$ if $\left\vert
\Delta\varphi^{P[AP]}\right\vert \gtrsim T_{1}^{s}+T_{2}^{s[\overline{s}]}$,
which we think possible in practice by using e.g. ferromagnetic insulators to
make the contacts between the leads and the wire (see above paragraph). Figure
\ref{Figure3}, left panel, shows the conductances $G^{P}$, $G^{AP}$ and the
magnetoresistance $MR=(G^{P}-G^{AP})/(G^{P}+G^{AP})$ for a device with weakly
transmitting barriers and $B=0$, in the absence of SDIPS (blue dashed lines)
or with a strong SDIPS such that $\varphi_{1}^{s}=\varphi_{2}^{s}$ for
$s\in\{u,d\}$ (red full lines). For convenience we plot the physical
quantities as a function of $\delta_{0}=L(k_{Fw}+[e\alpha V_{g}/\hbar
v_{Fw}])$ instead of the gate voltage $V_{g}$. The conductance presents
resonances with a $\pi$-periodicity in $\delta_{0}$. As explained above, the
SDIPS produces a spin-splitting of these resonances. Interestingly, this
increases significantly the $MR$ of the device by shifting the conductance
peaks in the $P$ and $AP$ configurations. When $\varphi_{1}^{s}\neq\varphi
_{2}^{s}$, both the $G^{P}$ and $G^{AP}$ curves can be spin-split, thus the
$MR$ curve can become more complicated (not shown), but this property persists
as long as $\Delta\varphi_{1}$ and $\Delta\varphi_{2}$ remain larger than the
transmission probabilities. When the transmissions become too large, it is not
possible to resolve $\Delta\varphi^{c}$ anymore because the dwell time of
electrons on the wire decreases. Then, it is not possible to have a giant
magnetoresistance. However, even in this situation, the SDIPS can modify
qualitatively the $MR$ of the device. Indeed, when there is no SDIPS, from the
expression of $\mathbb{T}_{ss}^{c}$, the $MR$ oscillations are always
symmetric with $V_{g}$. Even a weak SDIPS can break this symmetry (see Fig.
\ref{Figure3}, right panel). The reason for this phenomenon is that although
the transmission peaks associated to spins $\uparrow$ and $\downarrow$ are
merged, these peaks have spin-dependent widths due to the polarizations
$P_{1(2)}$ of the transmissions. Thus, the position of the global maximum
corresponding to $E_{\uparrow}^{c,j}$ and $E_{\downarrow}^{c,j}$ is different
in the $P$ and $AP$ configurations. For completeness, we recall that this
limit allows to obtain $MR(V_{g})$ curves strikingly similar \cite{MRdef} to
the $MR(V_{g})$ curves shown in \cite{Sahoo}.

\begin{figure}[ptb]
\centering\includegraphics[width=0.45\linewidth]{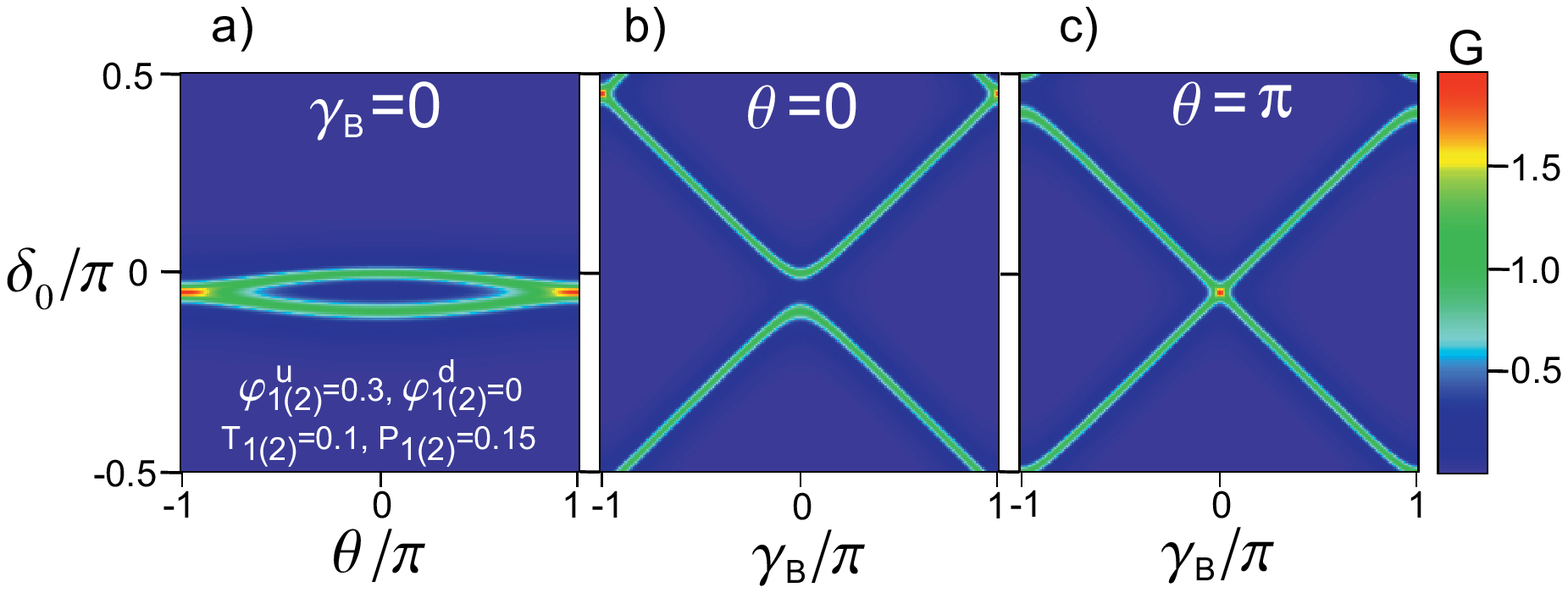}%
\quad\includegraphics
[width=.5\linewidth]{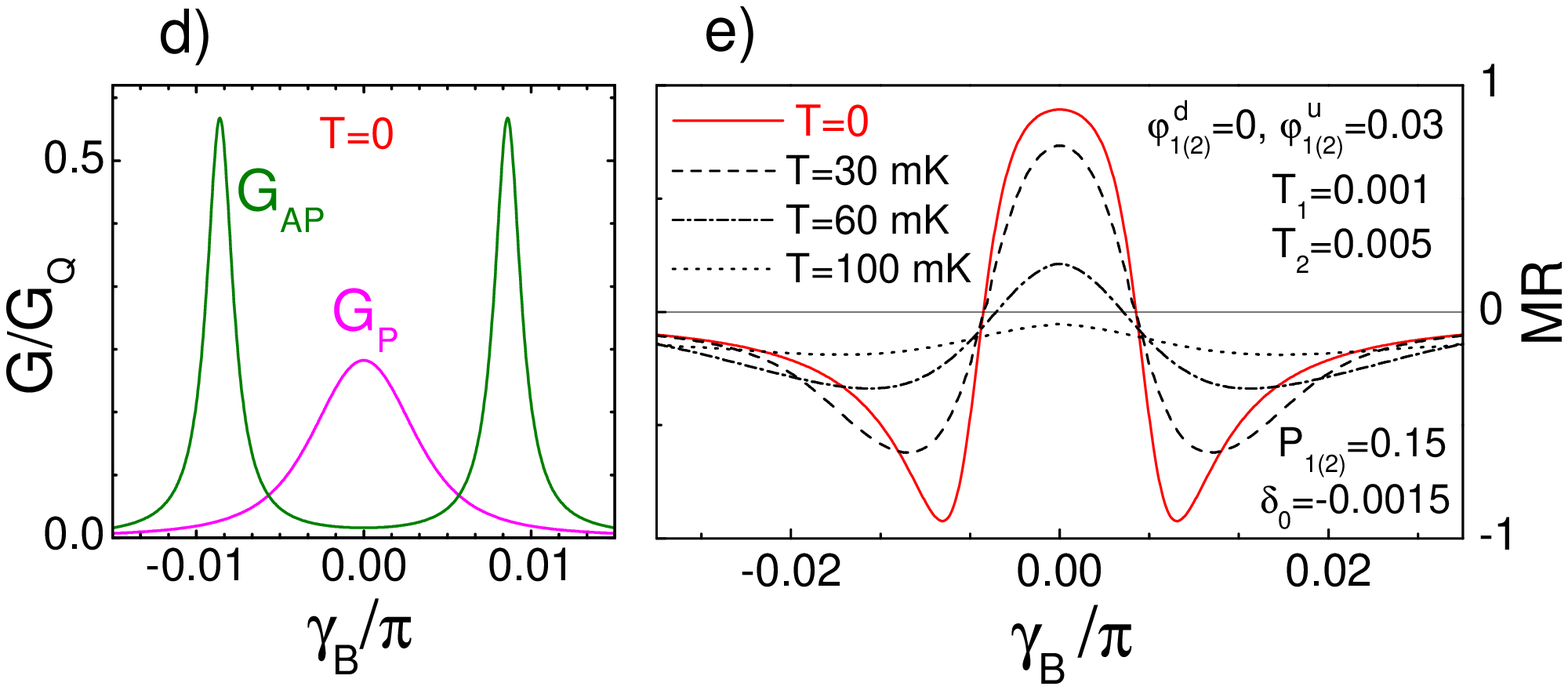}\caption{a,b: Color plots of the
conductance $G$ of the wire versus $\delta_{0}$ and $\theta$ for $\gamma
_{B}=0$ (plot a) or versus $\delta_{0}$ and $\gamma_{B}$ for $\theta=0 $ (plot
b) or $\theta=\pi$ (plot c). We used $T_{1}=T_{2}=0.1$, $P_{1}=P_{2}=0.15$,
$\varphi_{1}^{u}=\varphi_{2}^{u}=0.3$ and $\varphi_{1}^{d}=\varphi_{2}^{d}=0$.
c,d,e: conductances $G^{P}=G(\theta=0)$ and $G^{AP}$ $=G(\theta=\pi)$ (plot d)
and magnetoresistance $MR$ (plot e) as a function of $\gamma_{B}$. We used
$T_{1}=0.001$, $T_{2}=0.005$, $P_{1}=P_{2}=0.15$, $\delta_{0}=-0.0015$,
$\varphi_{1}^{d}=\varphi_{2}^{d}=0$, and $\varphi_{1}^{u}=\varphi_{2}%
^{u}=0.03$. All the data are shown for $T=0$ expect the magnetoresistance
which is shown also for finite temperatures. Finite temperatures curves are
plot for a wire with length $L=500~\mathrm{nm}$ and Fermi energy $v_{Fw}%
\sim8~10^{5}~\mathrm{m.s}^{-1}$. The $MR$ changes sign abruptly at a low value
of $\gamma_{B}$ because the resonances in $G^{AP}(\gamma_{B})$ and
$G^{P}(\gamma_{B})$ are shifted for the value of $\delta_{0}$ considered (this
configuration is possible thanks to the spin-dependent resonance pattern, as
can be understood from the color plots). The effect persists up to 90 mK for
the parameters chosen here. In this figure, we used $\varphi_{1}^{s}%
=\varphi_{2}^{s}$, thus there is no spin-splitting of the resonant energy at
$(\delta_{0}=0,\theta=\pi)$. However, we have checked that the field effect
shown in the bottom right panel can persist at $\varphi_{1}^{s}\neq\varphi
_{2}^{s}$. }%
\label{Figure4}%
\end{figure}

We now study non-collinear configurations. When $\theta\neq0[\pi]$ and $B=0$,
one has, for $s\in\{\uparrow_{1},\downarrow_{1}\}$ and $r\in\{\uparrow
_{2},\downarrow_{2}\}$%
\begin{equation}
\mathbb{T}_{sr}=\frac{A_{sr}^{P}}{\left\vert \mathbb{\beta}_{sr}^{P}%
\cos(\theta_{sr}/2)\left(  1+\gamma_{\phi}^{\kappa_{s}^{1}\kappa_{r}^{2}}%
\tan^{2}\left(  \theta_{sr}/2\right)  \right)  \right\vert ^{2}}
\label{TransTot}%
\end{equation}
with $\theta_{sr}=\widehat{(s,r)}$ and $\gamma_{\phi}=\mathbb{\beta}%
_{\uparrow_{1}\downarrow_{2}}^{P}\mathbb{\beta}_{\downarrow_{1}\uparrow_{2}%
}^{P}/\mathbb{\beta}_{\uparrow_{1}\uparrow_{2}}^{P}\mathbb{\beta}%
_{\downarrow_{1}\downarrow_{2}}^{P}$. Figure \ref{Figure4}, a, illustrates
that the spin-splitting in $G(\delta_{0})$ goes continuously from $\left\vert
\Delta\varphi^{P}\right\vert /2$ to $\left\vert \Delta\varphi^{AP}\right\vert
/2$ when $\theta$ goes from $0$ to $\pi$. This can be used to tune the
spin-splitting of the wire electronic spectrum. In the case $c\in\{P,AP\}$ and
$\theta_{B}=\pi/2$, one has, for $(s,r)\in\{\uparrow_{1},\downarrow_{1}\}^{2}%
$, an expression analogue to Eq.~(\ref{TransTot}), with $\theta_{sr}$ replaced
by $\gamma_{B}$, $\mathbb{\beta}_{sr}^{P}$ replaced by $\widetilde
{\mathbb{\beta}}_{sr}^{c}$ with $\widetilde{\mathbb{\beta}}_{sr}%
^{P[AP]}=1-\kappa_{s}^{1}\kappa_{r}^{1}B_{sr}^{P[AP]}e^{i(\varphi_{1}%
^{s}+\varphi_{2}^{r(\overline{r})}+2\delta)}$, $\gamma_{\phi}$ replaced by
$\widetilde{\gamma}_{\phi}^{c}=\widetilde{\mathbb{\beta}}_{\uparrow
_{1}\downarrow_{2}}^{c}\widetilde{\mathbb{\beta}}_{\downarrow_{1}\uparrow_{2}%
}^{c}/\widetilde{\mathbb{\beta}}_{\uparrow_{1}\uparrow_{2}}^{c}\widetilde
{\mathbb{\beta}}_{\downarrow_{1}\downarrow_{2}}^{c}$ and $\kappa_{r}^{2}$
replaced by $\kappa_{r}^{1}$. The signs $\kappa_{s(r)}^{1}$ in $\widetilde
{\mathbb{\beta}}_{sr}^{c}$ account for the $\pi$ phase shift acquired by a
spin $1/2 $ when the magnetic field makes this spin precess by $2\pi$. Due to
these signs, the dependence of $G$ on $\gamma_{B}$ can be very different from
its dependence on $\theta$. For instance, in Fig. \ref{Figure4}, a, b and c,
$G(\delta_{0},\gamma_{B}=0)$ shows resonances close to $\delta_{0}=0[\pi]$ for
any value of $\theta$ whereas the positions of the resonances in
$G^{P(AP)}(\delta_{0})$ strongly vary with $\gamma_{B}$, with avoided
crossings at $\delta_{0}\sim0[\pi]$ in the $P$ configuration and $\delta
_{0}\sim\pi/2[\pi]$ in the $AP$ configuration. For low transmissions, it is
possible to find a value of $\delta_{0}$ such that the conductances $G^{P}$
and $G^{AP}$ versus $\gamma_{B}$ display distinct resonances close to
$\gamma_{B}=0$ (Fig. \ref{Figure4}, d). This allows to obtain a giant
magnetoresistance whose sign can be switched by applying a magnetic field with
$\gamma_{B}\ll1$ (Fig. \ref{Figure4}, e). For instance, for $T_{1}=0.001$,
$T_{2}=0.005$, $P_{n}=0.15$, $\Delta\varphi_{n}=0.03$, $\delta_{0}=-0.0015$,
$L=500~\mathrm{nm}$ and $v_{Fw}\sim8~10^{5}~\mathrm{m.s}^{-1}$ \cite{bockrath}%
, one has $MR\sim+89\%$ at $B=0$ and $MR\sim-92\% $ at $B=250~\mathrm{mT}$
(Fig. \ref{Figure4}). This small value of magnetic field is particularly
interesting since in practice, it is difficult to keep $\vec{p}_{1}$ and
$\vec{p}_{2}$ perpendicular to $\vec{B}$ when $B$ becomes larger than
typically $1~\mathrm{T}$.

We now briefly address the effect of finite temperature $T$, which starts to
modify the behavior of the circuit when it becomes comparable to the wire
energy-level spacing $\hbar v_{Fw}/2L$ times the transmission probabilities
(see the above Breit-Wigner formula). The switching of the $MR$ sign with
$V_{g}$ described in Fig. \ref{Figure3}, left, is relatively robust to
temperature since it is still obtainable at 1 K for the wire parameters
considered in the previous paragraph (not shown). For Fig. \ref{Figure4}, e,
having a switching of the $MR$ sign with a low magnetic field requires to have
lower temperatures due the low values of transmission probabilities necessary.
However, this effect should be obtainable in practice since it persists up to
90 mK for the wire considered here (see Fig. \ref{Figure4}).

So far, we have disregarded the gate-dependence of the scattering matrices
$S_{1(2)}$. This is correct if the variations of $e\alpha V_{g}$ are
negligible compared to the characteristic energy scales defining the interface
scattering potentials. However, the opposite situation can occur. As an
example, we consider a wire with $v_{Fw}\sim8~10^{5}~\mathrm{m.s}^{-1} $ and
$L=500~\mathrm{nm}$, connected to two barriers like that described by the full
lines in Fig. \ref{Figure1}, right, with $T_{n}\sim0.1 $. The oscillation
period in $G(V_{g})$ is $T_{g}=\hbar v_{F}\pi/e\alpha L$. Starting from
$V_{g}=0$ for which $\Delta\varphi^{P}\sim0.4$, $\Delta\varphi^{P}$ varies by
only $0.15\%$ when $V_{g}$ changes by $2T_{g}$. Thus, on this scale, the SDIPS
can be considered as constant with $V_{g}$ and the previous approach is
correct. On larger scales, the periodicity of the $G(V_{g})$ curves is broken
and the SDIPS-induced spin-splitting of the resonant energies can be tuned
with $V_{g}$. For instance, a shift of $V_{g}$ by $70T_{g}$ makes
$\Delta\varphi^{P}$ vary by $\sim5\%$, i.e., $\sim100$ mT in terms of
effective magnetic field \cite{commentwire}. Thus, the effective field
produced by the SDIPS can be gate-dependent. The most simple consequence of
this feature is that the SDIPS-induced spin-splitting of the resonant energies
depends on the resonance index $j$. The gate-dependence of the SDIPS effective
field could be used for controlling the spin dynamics.

Before concluding, we note that our work has been used very recently by
\cite{morpurgo} for fitting $MR(V_{g})$ data obtained in a single wall
nanotube connected to ferromagnetic leads, in a regime in which Coulomb
blockade is absent. This provides further proof in favor of the relevance of
our approach, regarding single wall carbon nanotubes at least. These authors
have assumed to have no SDIPS, but considering the strong values of $T_{n}$ in
this experiment, the SDIPS is expected to cause only weak asymmetries in the
$MR(V_{g})$ curves, not resolvable in the actual experiment.

In summary, we have studied the effects of the spin-dependence of interfacial
phase shifts (SDIPS) on the linear conductance of a ballistic one-dimensional
quantum wire connected to two ferromagnetic leads. The SDIPS generates a
spin-splitting of the wire energy spectrum which is tunable with the gate
voltage and the angle between the ferromagnetic polarizations. This can lead
in particular to a giant magnetoresistance effect with a sign tunable with the
gate voltage and the magnetic field. These properties could be exploited for
manipulating spins in the context of spin electronics or quantum computing.

This work was supported by the RTN Spintronics, the Swiss NSF, the RTN DIENOW,
and the NCCR Nanoscience.


\begin{thebibliography}{99}
\bibitem{Prinz}G. Prinz, Science \textbf{282}, 1660 (1998).

\bibitem{Datta}S. Datta and B. Das, Appl. Phys. Lett. \textbf{56}, 665 (1990).

\bibitem{Schapers}T. Sch\"{a}pers \textit{et al.}, Phys. Rev. B \textbf{64},
125314 (2000); S. Krompiewski \textit{et al.}, PRB \textbf{69}, 155423 (2004).

\bibitem{LossDiVicenzo}D. Loss and D.P. DiVincenzo, Phys. Rev. A \textbf{57},
120 (1998).

\bibitem{Zutic}I. Zutic \textit{et al.}, Rev. Mod. Phys. \textbf{76}, 323 (2004).

\bibitem{FNF}A. Brataas, Yu. V. Nazarov, and G. E. W. Bauer, Phys. Rev.
Lett.\textbf{11}, 2481 (2000), D. Huertas-Hernando \textit{et al.}, Phys. Rev.
B \textbf{62}, 5700 (2000). A. Brataas \textit{et al.}, Eur. Phys. J. B
\textbf{22}, 99 (2001).

\bibitem{Luttinger}L. Balents and R. Egger, Phys. Rev. Lett. \textbf{85}, 3454
(2000); Phys. Rev. B \textbf{64}, 035310 (2001).

\bibitem{Wetzels}W. Wetzels, G. E. W. Bauer, and M. Grifoni, Phys. Rev. B
\textbf{72}, 020407 (R) (2005).

\bibitem{Tokuyasu}T. Tokuyasu \textit{et al.}, Phys. Rev. B \textbf{38}, 8823 (1988).

\bibitem{otherBC}A. Millis \textit{et al.}, Phys. Rev. B \textbf{38}, 4504
(1988); M. Fogelstr\"{o}m, \textit{ibid.} \textbf{62}, 11812 (2000); J.C.
Cuevas \textit{et al.} \textit{ibid.} \textbf{64}, 104502 (2001); N. M.
Chtchelkatchev \textit{et al.}, JETP Lett. \textbf{6}, 323 (2001); D.
Huertas-Hernando \textit{et al.}, Phys. Rev. Lett. \textbf{88}, 047003 (2002);
J. Kopu \textit{et al.}, \textit{ibid.} \textbf{69}, 094501 (2004); E. Zhao
\textit{et al.}, \textit{ibid.} \textbf{70}, 134510 (2004).

\bibitem{SF2005}A. Cottet and W. Belzig, Phys. Rev. B \textbf{72}, 180503 (2005).

\bibitem{Tedrow}P. M. Tedrow \textit{et al.}, Phys. Rev. Lett. \textbf{56},
1746 (1986).

\bibitem{Takis}T. Kontos \textit{et al.}, Phys. Rev. Lett. \textbf{86}, 304
(2001), \textit{ibid.} \textbf{89}, 137007 (2002).

\bibitem{bockrath}W. Liang \textit{et al.}, Nature \textbf{411}, 665 (2001).

\bibitem{Sahoo}S. Sahoo \textit{et al.}, Nature Phys. \textbf{2}, 99 (2005).

\bibitem{Veillette}M. Y. Veillette \textit{et al.}, Phys. Rev. B \textbf{69},
075319 (2004).

\bibitem{Blanter}Ya. M. Blanter and M. B\"{u}ttiker, Phys. Rep. \textbf{336},
1 (2000).

\bibitem{Cahay}M. Cahay \textit{et al.}, Phys. Rev. B \textbf{37}, 10125 (1988).

\bibitem{MI}X. Hao \textit{et al.}, Phys. Rev. B \textbf{42} 8235 (1990).

\bibitem{MRdef}Ref. \cite{Sahoo} shows data in terms of $MR^{\prime}%
=(G^{P}-G^{AP})/G^{AP}$, but for low polarizations, one has $MR^{\prime}%
\sim2MR$. Note that a more quantitative interpretation of these data would
require to take into account Coulomb blockade effects observable in this
experiment, but we expect that the SDIPS-induced spin-splitting persist in
this situation.

\bibitem{BTK}G.E. Blonder \textit{et al.}, Phys. Rev. B \textbf{25}, 4515 (1982).

\bibitem{commentwire}This can be done while fulfilling the assumption $e\alpha
V_{g}\ll E_{Fw}$ made in the article.

\bibitem{morpurgo}H.T. Man \textit{et al.}, cond-mat/0512505.
\end{thebibliography}
\end{document}